\documentclass[journal=jacsat,manuscript=article]{achemso}

\usepackage[version=3]{mhchem} 
\usepackage{amsmath}
\usepackage{amssymb}
\usepackage{subfigure}
\usepackage{xcolor}

\author{Selma Moqvist}
\affiliation
{Department of Computer Science and Engineering,
Chalmers University of Technology and University of Gothenburg,
SE-41296 Gothenburg, Sweden}
\author{Weilong Chen}
\affiliation
{Department of Computer Science and Engineering,
Chalmers University of Technology and University of Gothenburg,
SE-41296 Gothenburg, Sweden}
\author{Mathias Schreiner}
\affiliation
{Department of Computer Science and Engineering,
Chalmers University of Technology and University of Gothenburg,
SE-41296 Gothenburg, Sweden}
\author{Feliks N\"uske}
\affiliation
{Max-Planck-Institute for Dynamics of Complex Technical Systems, Magdeburg, Germany}

\author{Simon Olsson}
\email{simonols@chalmers.se}
\affiliation
{Department of Computer Science and Engineering,
Chalmers University of Technology and University of Gothenburg,
SE-41296 Gothenburg, Sweden}

\title[Thermodynamic Interpolation]
  {Thermodynamic Interpolation: A generative approach to molecular thermodynamics and kinetics}

\abbreviations{Machine Learning, Deep Generative Models, Molecular Simulations, Message Passing Neural Networks, Stochastic Interpolants}
\keywords{American Chemical Society, \LaTeX}

\begin{document}

\begin{tocentry}





\end{tocentry}

\begin{abstract}
Using normalizing flows and reweighting, Boltzmann Generators enable equilibrium sampling from a Boltzmann distribution, defined by an energy function and thermodynamic state. In this work, we introduce Thermodynamic Interpolation (TI), which allows for generating sampling statistics in a temperature-controllable way. We introduce TI flavors that work directly in the ambient configurational space, mapping between different thermodynamic states or through a latent, normally distributed reference state. Our ambient-space approach allows for the specification of arbitrary target temperatures, ensuring generalizability within the temperature range of the training set and demonstrating the potential for extrapolation beyond it. We validate the effectiveness of TI on model systems that exhibit metastability and non-trivial temperature dependencies. Finally, we demonstrate how to combine TI-based sampling to estimate free energy differences through various free energy perturbation methods and provide corresponding approximated kinetic rates estimated through generator extended dynamic mode decomposition (gEDMD).
\end{abstract}

\section{Introduction}
Computing molecular properties and observables, such as free energies, is of great interest in numerous scientific and engineering applications. In statistical mechanics, we can express many of these observables directly through the partition function, or normalizing constant of statistical distribution over microscopic configurations of a molecular system \cite{chandler}. Such a statistical distribution, or {\it ensemble} is defined by the macroscopic control variables, such as temperature, volume, pressure, and chemical potential, which are kept constant. The canonical Boltzmann distribution, $\mu(\mathbf{x})=\mathcal{Z}^{-1}\exp(-E(\mathbf{x})/kT)$ is for example characterized by constant temperature ($T$), volume ($V$), and particle number ($N$). 

Due to the high-dimensionality of molecular systems, direct computation of the partition function is intractable. Instead, we rely on simulation strategies such as Markov Chain Monte Carlo (MCMC) or Molecular Dynamics (MD) to draw samples from the distribution \cite{understandingmd}. However, for molecular systems, the high-dimensional free energy landscapes lead to impractically long simulations, to ensure the generation of independent sampling statistics. 

Numerous enhanced sampling methods are available, all aiming to accelerate sampling. These methods, modify the statistical ensemble to ensure faster traversal between free energy basins, or couple multiple thermodynamic ensembles {\it replicas}, or a combination of the two --- albeit with the constraint, that it is possible to reweigh the generated samples back to the correct ensemble \cite{Ferrenberg_1989,0801.1426}. Some influential examples include meta dynamics \cite{Laio2002}, or conformational flooding \cite{Grubmller1995}, replica-exchange and parallel tempering, \cite{Marinari_1992} umbrella sampling \cite{Torrie1977}. Hénin {\it et al.} recently surveyed numerous other approaches \cite{Henin2022}. Machine learning is having a dramatic impact on these strategies, in particular in helping identify collective variables \cite{Chen_2018,Wang2019,Herringer_2023, M_Sultan_2017}, which is lowering the need for manual trial-and-error optimization, and transformations which lower the number of replicas \cite{skipping_re_ladder} needed to ensure effective simulation.

Current enhanced sampling methods allow us to compute stationary observables such as free energies. However, as they usually involve biasing the molecular dynamics, recovery of unbiased dynamics or kinetics, is only possible in certain situations \cite{Tiwary2013,Voter1997,Grubmller1995}. Kinetic modeling using Markov state models (MSM) \cite{SchuetteFischerHuisingaetal.1998, Prinz2011,Swope2004,Buchete2008,Husic2018}, Koopman operator approaches \cite{wu2020variational,Nske2019,KLUS2020132416}, dynamic graphical models \cite{Olsson2019, Hempel2022}, transition path sampling \cite{Dellago1998,Bolhuis2002,Jung2023,2312.05340} and deep learning infused approaches \cite{Mardt_2018, Chen2019, Mardt2022,deepgenmsm}, take an alternative approach, leveraging either long unbiased simulations, or massively parallel short simulations collected using adaptive sampling strategies \cite{Bowman2010,Doerr2014,Bogetti2019,Betz2019}. These approaches allow us to uncover unbiased dynamics and facilitate the calculation of unbiased stationary and dynamic observables, yet remain costly from a computational perspective.

Deep generative models enable the development of fundamentally new approaches to equilibrium sampling and sampling of stochastic dynamics \cite{ito,2410.10605,timewarp,Hsu2024, 2409.17808, 2410.09667}. An important example is Boltzmann generators (BG) \cite{noe2019boltzmann} where an invertible deep neural network model, is trained to transform samples from a simple distribution, e.g. a normal distribution, to a complicated distribution, e.g. a Boltzmann distribution. In practice, this is often implemented using a {\it normalizing flow} \cite{rezende2016variational,Tabak2010DENSITYEB,noe2019boltzmann}. Subsequent efforts to improve accuracy and transferability of BGs, leverage alternative neural networks, including diffusion models \cite{diez2024generation}, and continuous normalizing flows \cite{kohler2020equivariant,transbgs}, in a manner that account for molecular symmetries. A variation of this idea, called {`\it Boltzmann Emulators'} generate ensembles in a reduced configurational space, e.g. by parameterizing a generative model over torsion angles, and keeping bond-lengths and angles at idealized values, have also shown some success \cite{diez2024generation, 2206.01729, Zheng2024}. 

The connections between statistical mechanics, and deep generative models have further fueled a zoo new methods for sampling. Many of these are either inspired by perturbative methods or perturbative in nature including thermodynamic maps \cite{herron2023inferring}, which learn a coarse-grained multi-temperature model from replica-exchange data. Other examples include mapping between cheap reference potentials and expensive quantum mechanical models \cite{tfep_revisisted_rizzi}, computing free energies \cite{Wirnsberger_2020}, or decreasing the number of replicas in a replica exchange scheme \cite{skipping_re_ladder}. More recently, a denoising diffusion model was used to perform a learned thermodynamic integration between in ideal gas and a Lennard-Jones liquid \cite{máté2024neuralthermodynamicintegrationfree}. Finally, an early example proposed constructing normalizing flows that were steerable under temperature transformations \cite{dibak2022temperature}, allowing for some generalization across temperature. 

In this work, we propose Thermodynamic Interpolation (TI) as a method for generating samples across multiple thermodynamic states, through a learned map between different Boltzmann distributions. We present two different TI methods {\bf Ambient TI} and {\bf Latent TI}. Ambient TI  transforms between two thermodynamic states directly in the configurational space, while latent TI transform samples between thermodynamic ensembles through a normally distributed latent space distribution. We implement both TI approaches using new simulation-free training schemes developed for continuous normalizing flows \cite{albergo2023stochastic}, and consequently, we can both transform samples between distributions and compute their change in log probabilities. Using temperature transformations as an example, we demonstrate that our ambient TI approach enables transformations between ensembles at different temperatures in configuration space. Our latent TI models similarly enable transformations between ensembles albeit through a shared latent space distribution, or reference state. Further, since the latent TI is implemented using a temperature conditioned BG we can generate samples at multiple different temperatures on-demand. We find both methods can be trained very efficiently with limited simulation data, and information is shared between thermodynamically similar ensembles. We evaluate our approach on a one-dimensional double-well model system and then scale it to MD simulations \cite{diez2024generation} of two molecules from the QM9 dataset \cite{ramakrishnan2014quantum}: N-Methylformamide (N-Me) and 3-propan-2-ylhex-1-yne (3p2y1y). For all models, we achieve high sampling efficiency, even outside the training data, allowing us to estimate equilibrium properties such as free energy differences and dynamic properties like kinetic transition rates at temperatures not encountered during training. We thus present a framework that enables flexible transformations between arbitrary thermodynamic states, along with access to corresponding probabilities. As a result, we believe our TI approach offers a robust and generalizable tool for accurately predicting thermodynamic and kinetic properties across a wide range of thermodynamic states.

\section{Methods}
\subsection{Estimation of Free Energies}
Free energy perturbation (FEP)\cite{Zwanzig1954HighTemperatureEO} is a method to compute free energy difference $\Delta F_{AB}$ between two thermodynamic states, $A$ and $B$, through the identity
\begin{equation}\label{eqn: fep_def}
    \mathbb{E}_{\mathbf{x}_A}\left[\exp(-(E_B(\mathbf{x}_A)-E_A(\mathbf{x}_A)))\right]=\exp(-\Delta F_{AB}),
\end{equation}
where $E_A$ and $E_B$ are unit-less potential energy functions, and averaging is done for the Boltzmann distribution of state $A$, 
\begin{equation}
    \mathbf{\mu}_A(\mathbf{x}) = \mathcal{Z}_A^{-1}\exp(-E_A(\mathbf{x})).
\end{equation}
Unfortunately, this free energy estimator is inefficient if the overlap between states $A$ and $B$, is low.

To alleviate this, Jarzynski introduced targeted FEP (TFEP) \cite{PhysRevE.65.046122} which augments the FEP by introducing a map $f_{AB}:\, \mathbf{x}_A \rightarrow \mathbf{x}_B$ aims to transform the state $A$ to another state closer to $B$. If such map can be found, and is differentiable and invertible, then the free energy difference between $A$ and $B$ can be estimated only given samples $\mathbf{x}_A\sim \mu_A$ by computing the average
\begin{equation}
\mathbb{E}_{\mathbf{x}_A}\left[\exp(-\varphi(\mathbf{x}_A))\right] = \exp(-\Delta F^\mathrm{(TFEP)}),
\end{equation}
where 
\begin{equation}\label{eqn:phi_tfep_def}
    \varphi(\mathbf{x}_A) \triangleq E_B\left(f_{AB}(\mathbf{x}_A)\right) - E_A(\mathbf{x}_A) - \log \det \left(J_{f_{AB}}(\mathbf{x}_A)\right).
\end{equation}
While this approach enjoys more favorable convergence properties, determining the map $f_{AB}$ for a given application is not trivial. With the emergence of deep generative models, several approaches have been proposed to learn such maps for a variety of interesting applications  \cite{Wirnsberger_2020,herron2023inferring,tfep_revisisted_rizzi, skipping_re_ladder, Willow2023}. 

An alternative strategy to free energy difference estimation is based on Boltzmann generators (BG) \cite{noe2019boltzmann}. BGs are implemented with a normalizing flow, which is trained to transform the normally distributed latent space $\mathbf{z}\sim\mathcal{N}(0,\,\mathrm{Id})\triangleq\mu_Z$ into molecular conformations $f_{ZA}^{(\theta)}(\mathbf{z})\sim\rho_A\approx\mu_A$. In their paper No\'e and co-workers \cite{noe2019boltzmann}, show that the free energy difference between two meta-stable states $A$ and $B$ can be computed using two independent BGs, transforming from a latent state $Z$ one into $A$ and the other into $B$, using the expression
\begin{equation}\label{eqn: dF_two_BGs_JKL}
    \Delta F^\mathrm{(BG)} = \mathbb{E}_\mathbf{z}\left[\mathcal{L}_\mathrm{KL}^{(\mathrm{BG},\,B)}(\mathbf{z})\right] - \mathbb{E}_\mathbf{z}\left[\mathcal{L}_\mathrm{KL}^{(\mathrm{BG},\,A)}(\mathbf{z})\right],
\end{equation}
where the BG loss function $\mathcal{L}_\mathrm{KL}^\mathrm{(BG, X)}$ equal to the free energy of state $X$ up to a constant. 
Using the definition of BG Kullback-Leibler loss we can rewrite the estimator as
\begin{equation}\label{eqn: bg_general_dF_main}
    \begin{split}
        \Delta F^\mathrm{(BG)} &= \mathbb{E}_\mathbf{z}\left[E_B(f_{ZB}(\mathbf{z})) - E_A(f_{ZA}(\mathbf{z}))+ \log\left\lvert\det J_{f_{ZA}}(\mathbf{z})\right\rvert-\log\left\lvert\det J_{f_{ZB}}(\mathbf{z})\right\rvert\right]\\
        &\triangleq\mathbb{E}_\mathbf{z}\left[\psi(\mathbf{z})\right].
    \end{split}
\end{equation}
As we show in the Supporting information (free energy perturbation methods), it follows from the invertibility of the BG map that $\psi$ is equivalent to the function (eq.~\ref{eqn:phi_tfep_def}). This further suggests that
\begin{equation}\label{eqn:tfep_for_bg}
    \mathbb{E}_\mathbf{z}\left[\exp(-\psi(\mathbf{z}))\right]=\frac{\mathcal{Z}_B}{\mathcal{Z}_A}\triangleq\exp(-\Delta F^\mathrm{(TFEP)}).
\end{equation}
As implied by Equations \eqref{eqn: bg_general_dF_main} and \eqref{eqn:tfep_for_bg}, it is not necessary to have an explicit map between the two thermodynamic states, but rather it is sufficient with a map that transforms some initial distribution into the two target states $A$ and $B$ separately. Similarly to previous work \cite{Jarzynski_1997,tfep_revisisted_rizzi}, we can apply Jensen's inequality to obtain a relationship between the BG and TFEP estimators, given as 
\begin{equation}\label{eqn: estimator_relation}
    \Delta F^\mathrm{(BG)} \geq\Delta F^\mathrm{(TFEP)}.
\end{equation}
The inequality~\eqref{eqn: estimator_relation} can also be seen as an expression of the Donsker-Varadhan (DV) variational principle for importance sampling of exponential expectations~\cite{deuschel2001large,hartmann2017variational}. Here, we consider changes of probability measure induced by the push-forward under the invertible transformation $f_{ZB}\circ f_{ZA}^{-1}$. The DV principle states that~\eqref{eqn: bg_general_dF_main} is an upper bound for the exponential expectation~\eqref{eqn:tfep_for_bg}. In addition, the variance of~\eqref{eqn: bg_general_dF_main} is decreasing with increasing right-hand side, leading to a (theoretical) one-shot estimator if equality is attained in~\eqref{eqn: estimator_relation}. As such, the TFEP-estimator serves as a lower bound on the BG-estimator. A more in-depth derivation and discussion of this can be found in the Supporting information (free energy perturbation methods).

\subsection{Continuous Normalizing Flows}
Normalizing flows is a class of deep generative models, where we learn a map $f^{(\theta)}: \Omega_0 \to \Omega_1$, parameterized by $\theta$ where $\Omega_0,\,\Omega_1\subset\mathbb{R}^d$ and $d \in \mathbb{N}$, that transforms an initial distribution $\rho_0: \Omega_0 \to \mathbb{R}_+$ into a target distribution $\rho_1: \Omega_1 \to \mathbb{R}_+$. To obtain sample probabilities, the map must be both smooth and invertible, a {\it diffeomorphism}. In general, the map can be between any two distributions, for example Boltzmann distributions at different temperatures. 

A diffeomorphic map can be constructed as either a composition of several smooth and invertible partial transformations \cite{dinh2017densityestimationusingreal} or as the solution to an initial value problem, known as a continuous normalizing flow (CNF)\cite{chen2018neural}. In other words, we learn the velocity field $\mathbf{b}^{(\theta)}:\,\left[0,\,1\right]\times\mathbb{R}^d\to\mathbb{R}^d$ such that ordinary differential equation (ODE)
\begin{equation}\label{eqn: per_sample_cnf_de}
    \frac{\mathrm{d}\mathbf{x}(t)}{\mathrm{d}t} = \mathbf{b}^{(\theta)}(t,\,\mathbf{x}(t)),\quad \mathbf{x}(0) = \mathbf{x}_0,
\end{equation}
with initial condition $\mathbf{x}_0 \sim \rho_0$, approximates $\rho_1$ when integrated in time from $0$ to $1$. We call the resulting map $f^{(\theta)}_{01}$.  The ODE coupled with the initial distribution $\rho_0$ gives rise to a time-dependent probability density $\rho(t,\mathbf{x}(t))$ described by the continuity equation,
\begin{equation}\label{eqn: cont_eqn}
    \frac{\partial\rho }{\partial t} + \nabla \cdot (\rho \mathbf{b}^{(\theta)}) = 0,\quad \rho(0,\,\mathbf{x}(0)) = \rho_0(\mathbf{x}_0).
\end{equation}
We can solve both equations jointly to get both the transformed sample $f^{(\theta)}_{01}(\mathbf{x}_0) = \mathbf{x}_1 \sim \rho_1$ and its corresponding change in logarithmized probability or Jacobian determinant $\log \left\lvert\det J_{f^{(\theta)}_{01}}(\mathbf{x}_0)\right\rvert$, where the latter is obtained by integrating
\begin{equation}
    \log\rho(t,\mathbf{x}(t)) = \log\rho(0,\,\mathbf{x}(0)) - \underbrace{\int_0^1\nabla\cdot\mathbf{b}^{(\theta)}(s,\,\mathbf{x}(s))\mathrm{d}s}_{\log \left\lvert\det J_{f^{(\theta)}_{01}}(\mathbf{x}_0)\right\rvert },
\end{equation}
using an off-the-shelf numerical solver.

Here we use the Stochastic Interpolant (SI) framework to learn the velocity field $\mathbf{b}^{(\theta)}$\cite{albergo2023stochastic}. By interpolating between samples from the initial and target distributions in a stochastic manner, one can define a stochastic process 
\begin{equation}\label{eqn: stoch_interp_def}
    \mathbf{x}(t) = I(t,\,\mathbf{x}_0,\,\mathbf{x}_1) + \gamma(t)\mathbf{z},\quad t\in[0,\,1],\quad\mathbf{z}\sim\mathcal{N}(0,\,\mathrm{Id}),
\end{equation}
where the interpolant $\mathbf{x}(t)$ satisfies $\mathbf{x}(0) = \mathbf{x}_0$ and $\mathbf{x}(1) = \mathbf{x}_1$. The interpolant in Equation \eqref{eqn: stoch_interp_def} defines a path for moving samples from $\rho_0$ to $\rho_1$ in finite time, with the guarantee that at time $t=0$ the sample is distributed according to $\rho_0$, and at time $t=1$ the sample is distributed according to $\rho_1$. However, at intermediate times $t\in(0,\,1)$ the interpolant $\mathbf{x}(t)$ is completely characterized by the functions $I$ and $\gamma$. Since these can be chosen in any way that respects the SI boundary conditions\cite{albergo2023stochastic}, the velocity of the interpolant, $\dot{\mathbf{x}}(t)$, is easily obtained as long as one uses appropriate choices of $I$ and $\gamma$. A vector field $\mathbf{b}$ can be defined as the expected velocity of the interpolant (eq.~\ref{eqn: stoch_interp_def})
\begin{equation}\label{eqn: velocity_def}
\mathbf{b}(t,\,\mathbf{x}(t)) = \mathbb{E}\left[\Dot{\mathbf{x}}(t)\lvert\,\mathbf{x}(t)\right],
\end{equation}
where we note that $\mathbf{b}$ describes the velocity of the individual samples in Equation $\eqref{eqn: per_sample_cnf_de}$ and the density in Equation \eqref{eqn: cont_eqn}. In order to learn a parameterized version, $\mathbf{b}^{(\theta)}$, of the velocity $\mathbf{b}$, we minimize the regression-based objective
\begin{equation}\label{eq:loss}
    \mathcal{L}\left[\mathbf{b}^{(\theta)}\right] =\mathbb{E}_{\mathbf{x}_0,\mathbf{x}_1,t}\left[\frac{1}{2} \left\lvert\mathbf{b}^{(\theta)}(t,\,\mathbf{x}(t))\right\rvert^2-\left(\partial_tI(t,\,\mathbf{x}_0,\,\mathbf{x}_1) + \dot{\gamma}(t)\mathbf{z}\right)\cdot \mathbf{b}^{(\theta)}\left(t,\,\mathbf{x}(t)\right)\right]
\end{equation}
where times $t$ are drawn as $t\sim\mathcal{U}[0,\,1]$.

\subsection{SE(3)-Equivariant Message Passing Neural Networks}
The energy of a closed molecular system is symmetric under $\mathrm{E}(3)$ group action. These symmetries imply that models of the potential energy and the corresponding Boltzmann distribution should be {\it invariant} to global 3D rotations, translations, and inversion \cite{1802.08219, kohler2020equivariant}. Consequently, when we learn functions to approximate potential energies or Boltzmann distributions from data we can limit our hypothesis space to functions that satisfy these symmetries without incurring any approximation error \cite{2106.07148}. A consequence of this is improved data-efficiency and better generalization beyond the training set.

More formally, we call a function $f$ $G$-invariant if $f(T_g\mathbf{x})=f(\mathbf{x})$ and $G$-{\it equivariant} if $S_gf(\mathbf{x}) = f(T_g\mathbf{x})$, where $S_g$ and $T_g$ are linear representations of the group element $g\in G$. We can learn a $G$-invariant probability density model by combining a $G$-invariant distribution $\rho_0$ with a $H$-equivariant velocity field \cite{kohler2020equivariant}, where $H$ is a subgroup of $G$. In other words, applying equivariant perturbations to an invariant probability density will leave it invariant.  Practically, the neural network parameterizing our velocity field $\mathbf{b}^{(\theta)}$ should be constructed in a way that is equivariant with respect to the symmetries displayed by a molecular system.

The energy of a molecule generally exhibits reflection and permutation symmetry. However, as we here work with classical MD data, where chirality is usually constrained we follow previous work and instead of the $\mathrm{E}(3)$-group we thus consider the $\mathrm{SE}(3)$-group, which is comprised of only rotations and translations. A neural network architecture fulfilling this requirement is the ChiroPaiNN architecture, which we employ to build $\mathrm{SE}(3)$-equivariant flow models \cite{NEURIPS2023_7274ed90}. While the architecture we use here is equivariant under the permutation group, we do not exploit it in our experiments on molecules.

\subsection{Thermodynamic Interpolation}
\begin{figure}[h!]
    \centering
    \includegraphics[width=0.7\linewidth]{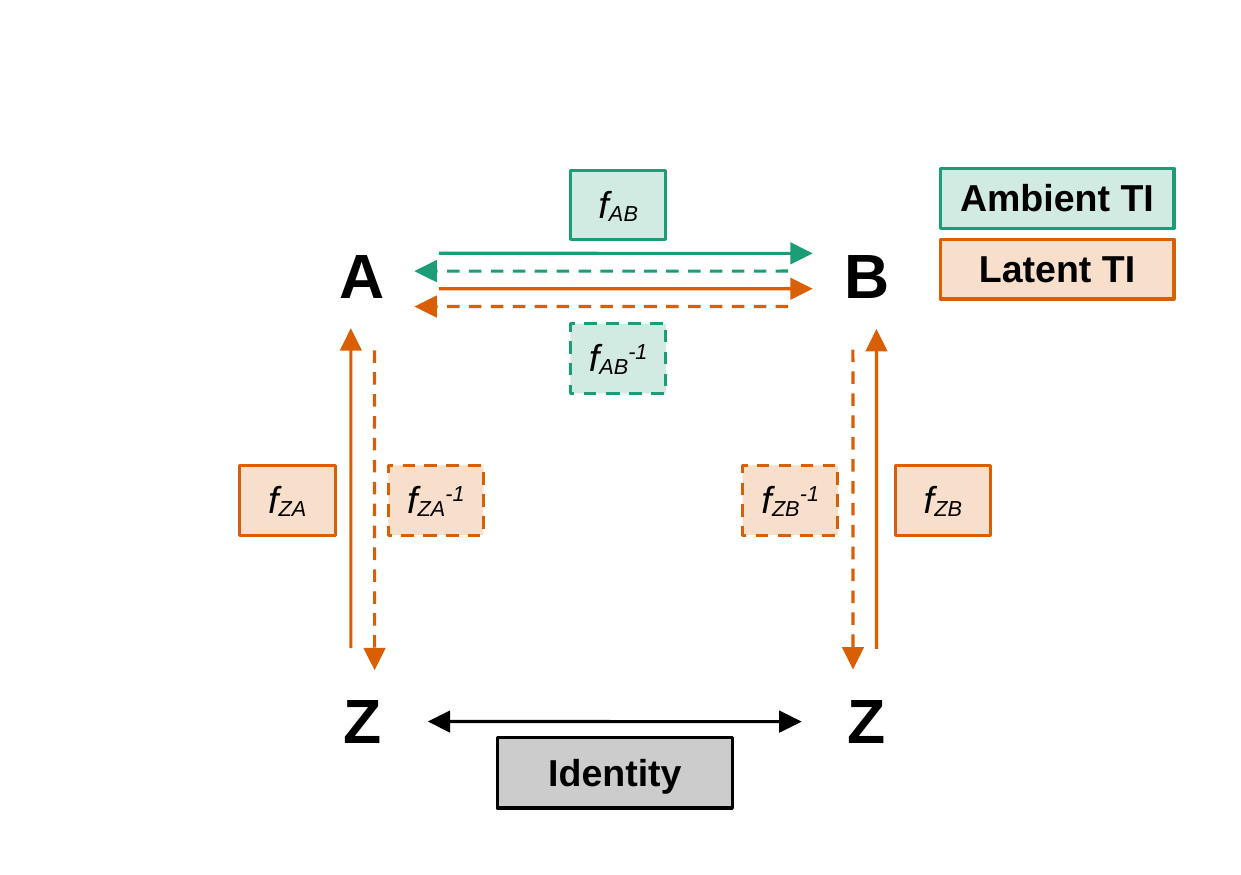}
    \caption{{\bf Illustration of the two thermodynamic interpolation approaches.} In the ambient case (green) one begins with Boltzmann distributed samples $\mathbf{x}_A\sim\mu_A$ at the thermodynamic state $A$. These are transformed the into samples $f_{AB}^{(\theta)}(\mathbf{x}_A)\sim\rho_B$ drawn from the surrogate distribution at state $B$, using the learned aTI map $f_{AB}^{(\theta)}$. Finally, when estimating observables, samples are weighed using importance weights $w^a$ so that they correspond to the thermodynamic state $B$. This is done to correct for the bias introduced by the surrogate. One can also first learn a lTI surrogate model to generate approximately Boltzmann distributed samples $f_{ZA}^{(\theta)}(\mathbf{z})\sim\rho_A$, compute the corresponding weights $w^l$ to weigh $f_{ZA}^{(\theta)}(\mathbf{z})$ into $\mathbf{x}_A$, and then transform $\mathbf{x}_A$ into $\mathbf{x}_B\sim\mu_B$ with aTI (orange) and the corresponding weights $w^a$.}
    \label{fig:ti_bg_thermo_relationship}
\end{figure}

In this work, we present thermodynamic interpolation (TI), an approach which builds on the idea of targeted free energy perturbation \cite{PhysRevE.65.046122} where a diffeomorphic map is used to transform between different thermodynamic states with the aim to compute free energy differences. Here, we learn such maps using two different approaches: latent and ambient TI (Figure \ref{fig:ti_bg_thermo_relationship})

\paragraph{Latent TI} transforms samples between thermodynamic states through a latent space equipped with a normal distribution.
We implement a latent TI (lTI) model using a temperature conditioned BG.
 Due to modeling errors, the BG will sample from an approximation, $\rho_A$, of the true Boltzmann distribution, $\mu_A$. To compute unbiased estimates we therefore need to weigh samples by their importance weights \cite{noe2019boltzmann}
\begin{equation}\label{eq:importance_wbg}
    w_\mathrm{BG}(\mathbf{z}) = \exp\left(-\frac{1}{kT_A}E(f_{ZA}^{(\theta)}(\mathbf{z})) - \log\mu_Z(\mathbf{z}) + \log\left\lvert\det J_{f_{ZA}^{(\theta)}}(\mathbf{z})\right\rvert\right).
\end{equation}
In lTI, we make the map temperature-conditioned $f_{ZA}^{(\theta)}(\cdot;T_A)$, in a similar spirit to recent work aiming to compute phase diagrams using normalizing flows \cite{schebek2024}.

To train our temperature conditioned map we use make of the {\it 'one-sided interpolant'}
\begin{equation}\label{eqn:interpolant_def_lTI}
    I(t,\,\mathbf{z},\,\mathbf{x}_A) = \left(1-t\right)\mathbf{z} + t\mathbf{x}_A = \mathbf{x}(t),
\end{equation}
linearly interpolating normally distributed noise $\mathbf{z}\sim\mu_Z$ with Boltzmann distributed configurations $\mathbf{x}_A\sim\mu_A$ at temperature $T_A$. By combining this interpolant, with the loss (eq.~\ref{eq:loss}), which further averages is over the thermodynamic states $A$, in our case, temperatures,
\begin{equation}
    \mathcal{L}[\mathbf{b}^{(\theta)}] = \mathbb{E}_{t,\mathbf{z},\mathbf{x}_A,A}\left[\frac{1}{2} \lvert \mathbf{b}^{(\theta)}(t,\,\mathbf{x}(t);T_A)\rvert^2-\partial_tI(t,\,\mathbf{z},\,\mathbf{x}_A) \cdot \mathbf{b}^{(\theta)}\left(t,\,\mathbf{x}(t);T_A\right)\right],
\end{equation}
we can learn a parameterized version $f^{(\theta)}_{ZA}$ of the lTI map $f_{ZA}$ which is conditioned on temperature. Then we can generate samples at an arbitrary temperature $T_A$ by sampling the latent space distribution and transforming them with the learned forward map $f_{ZA}^{(\theta)}$, and compute unbiased observables by weighing the samples by their importance weights (eq.~\ref{eq:importance_wbg}).


\paragraph{Ambient TI}
In many cases, two thermodynamic ensembles may be more similar to each other than to the reference state encoded in a latent space. Consequently, we propose ambient TI (aTI) as an approach to learn a direct map between thermodynamic states $f_{AB}$ in the configuration space. Compared to lTI, learning a direct map will avoid two step transformations. For example computing free energy changes between two thermodynamic ensembles using lTI would require sampling positions and their corresponding probabilities at two temperatures. With a direct map we could sample at one temperature and directly compute the free energy from the change in probability associated of the aTI map using these samples. Additionally, we can also combine an aTI map with a Parallel Tempering protocol, akin to previous work \cite{skipping_re_ladder}, to avoid having to simulate multiple replicas to ensure efficient sampling. As illustrated in Figure \ref{fig:ti_bg_thermo_relationship},  high-temperature (e.g. $A$) samples could then be mapped directly into lower temperatures (e.g. $B$) using aTI.

To train aTI maps $f_{AB}$ between thermodynamic states $A$ and $B$ we make use of the {\it`two-sided interpolant'} \cite{albergo2023stochastic}
\begin{equation}
    I(t,\,\mathbf{x}_A, \mathbf{x}_B, \mathbf{z}) = (1-t)\mathbf{x}_A + t\mathbf{x}_B + \gamma(t)\mathbf{z} = \mathbf{x}(t),
\end{equation}
where $\mathbf{z}\sim\mu_Z$, $\mathbf{x}_A\sim \mu_A$ and $\mathbf{x}_B\sim \mu_B$. Here, $\mu_A$ and $\mu_B$ represent Boltzmann distributions with identical energy functions $E$ but at different temperatures $T_A$ and $T_B$, corresponding to states $A$ and $B$ respectively. We detail the choice of $\gamma$, in the Supporting information (hyperparameters for ADW system and hyperparameters for molecular systems). As for the lTI, we can combine this interpolant, with the loss (eq.~\ref{eq:loss}), where we now take the expectation with respect to source and target state temperatures, samples drawn at both temperatures ($\mathbf{x}_A$ and $\mathbf{x}_B$), $t$ and $\mathbf{z}$,
\begin{equation}
    \mathcal{L}[\mathbf{b}^{(\theta)}] = \mathbb{E}_{t,\mathbf{z},\mathbf{x}_A,\mathbf{x}_B,A,B}\left[\frac{1}{2} \lvert \mathbf{b}^{(\theta)}(t,\,\mathbf{x}(t);T_A, T_B)\rvert^2-\partial_tI(t,\,\mathbf{x}_A, \mathbf{x}_B, \mathbf{z}) \cdot \mathbf{b}^{(\theta)}\left(t,\,\mathbf{x}(t);T_A, T_B\right)\right].
\end{equation}
Note, that our model of the velocity field $\mathbf{b}^{(\theta)}$ now depends on source and target temperatures.

When applying aTI, we can provide samples to the initial state in several ways. For example, we can run conventional, or enhanced sampling, MD simulations at $T_A$, or we can sample a surrogate model which generates sample from $\rho_A\approx\mu_A$. By applying the learned map we can then transform the samples to samples from the surrogate $\rho_B$, and we can recover unbiased estimates of observables by computing the importance weights
\begin{equation}\label{eqn: tbg_weights}
    w^a(\mathbf{x}_A,\,T_A,\,T_B) = \exp\left(-\frac{1}{kT_A}E\left(\mathbf{x}_A\right) - \frac{1}{kT_B}E\left(f_{AB}^{(\theta)}\left(\mathbf{x}_A\right)\right) + \log\lvert\det J_{f_{AB}^{(\theta)}}(\mathbf{x}_A)\rvert\right).
\end{equation}
When initial conditions are generated with a surrogate, we further need to account for the approximate nature of the surrogate, e.g. using the important weights if its a (temperature conditioned) BG $w_\mathrm{BG}$. We then use Equation \eqref{eqn: tbg_weights} to compute the aTI weights $w^a$.

\begin{figure}
    \centering
    \includegraphics[width=0.8\linewidth]{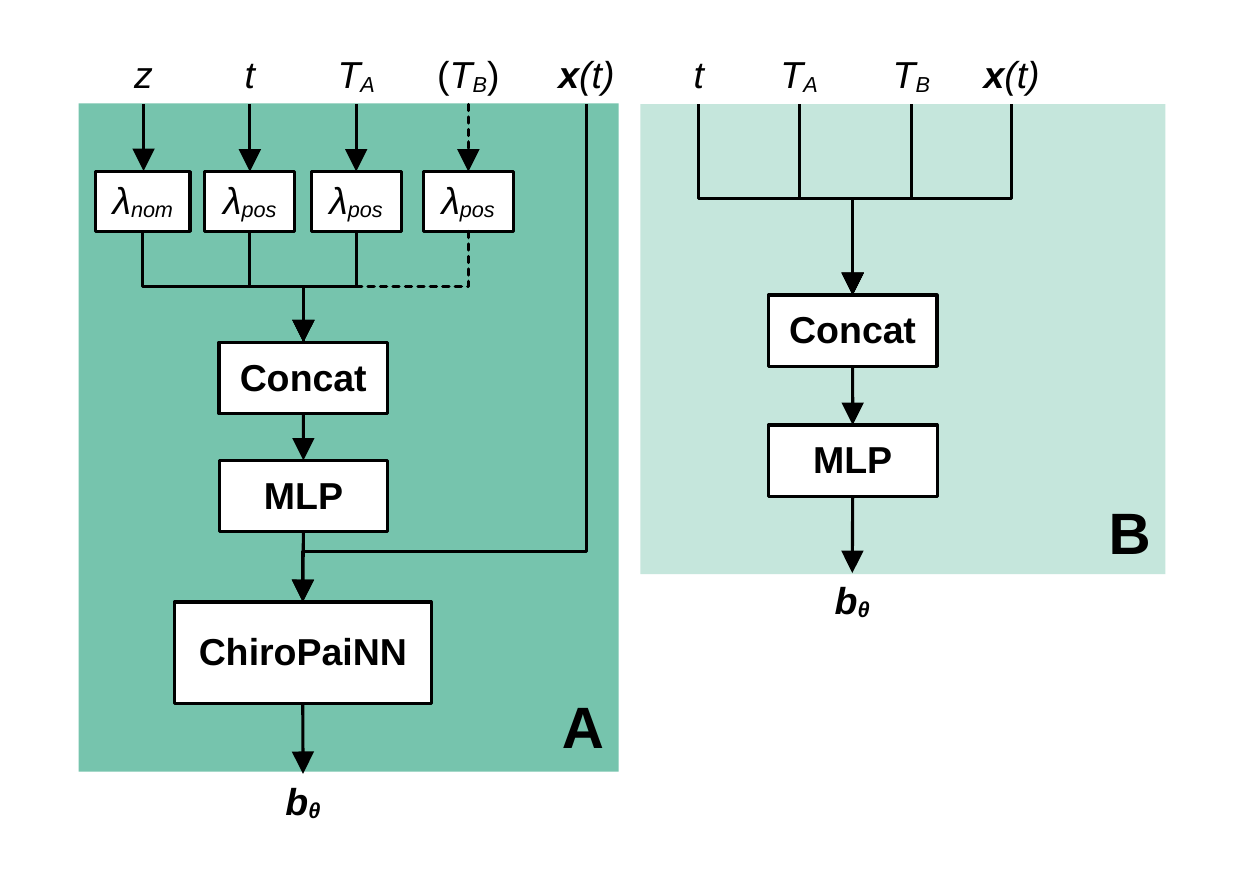}
    \caption{{\bf Thermodynamic Interpolant neural architecture.}\textbf{A} Latent and ambient TI for high-dimensional molecular-type systems. In the latent case we are only interested in sampling at a single temperature $T_A$ and do not input the second temperature $T_B$ to the cPaiNN model. In the ambient case, both temperatures $T_A$ and $T_B$ are included, along with atom numbers $z$. \textbf{B} Ambient TI for low-dimensional systems. In the lower-dimensional case we do not encode the temperatures or times into positional embeddings $\lambda_\mathrm{pos}$, and a simple MLP can be used instead of the ChiroPaiNN model.}
    \label{fig:tbg_and_ti_models}
\end{figure}

\paragraph{Implementation details} The learned velocity fields $\mathbf{b}^{(\theta)}$ depend on a range of inputs. We illustrate the information flow in Figure \ref{fig:tbg_and_ti_models}. For the molecular systems, we embed the temperature using the positional embeddings $\lambda_\mathrm{pos}(T)=\{\lambda_\mathrm{pos}^n(T)\}_{n=0}^N$ of dimension $N$, where
\begin{equation}\label{eqn: pos_encodings_def}
    \lambda_\mathrm{pos}^n(T)= 
    \begin{cases}
        \cos\left(\left(1 + \frac{n}{2}\right)\frac{\pi \hat{T}}{l_0}\right)\quad\text{for even }n\\
        \sin\left(\left(1 + \frac{n-1}{2}\right)\frac{\pi \hat{T}}{l_0}\right)\quad\text{for odd }n,
    \end{cases} 
\end{equation}
for normalized temperatures $\hat{T} = \frac{T - \Bar{T}}{T_\mathrm{max} - T_\mathrm{min}}$ and where $l_0$ is a hyperparameter. By changing $l_0$, embeddings of different temperatures can be made more or less similar to each other. We embed the interpolation time $t$ and atom numbers $z$ are positional and  nominal embeddings, following previous work \cite{NEURIPS2023_7274ed90}. For the asymmetric double-well system we use the temperatures directly. Specific details on hyperparameter for experiments are available in the Supporting information (Hyperparameters for ADW system and hyperparameters for molecular systems)

\subsection{Generator Extended Dynamic Mode Decomposition}
Modeling slow processes such as protein folding and ligand-binding/unbinding to a target protein is challenging as it relies on extensive unbiased MD simulation data. However, recent advances allow for data-driven estimation of the Koopman operator through extended dynamic mode decomposition (EDMD)\cite{Williams2015,klus_data-driven_2018}, which can be extended to the Koopman generator by using generator EDMD\cite{KLUS2020132416}. With statistical estimates of Koopman operators or generators, for example expressed in some feature space, we analyse slow processes through their eigenvectors and eigenvalues, akin to MSMs.

For lag time $t \geq 0$, the Koopman operator maps $\phi$ to the conditional expectation
\begin{equation}
     \mathcal{K}^t \phi(x)= \mathbb{E} \left[\phi\left(\mathbf{x}_t\right) \mid \mathbf{x}_0=x\right],   
\end{equation}
where $\mathbf{x}_t$ is a (stochastic) dynamical system in the space $\mathbb{X} = \mathbb{R} ^d$, and $\phi$ is a function of the 
 state space. By taking the time derivative of $\mathcal{K}^t \phi$ at $t=0$, we obtain the Koopman generator
\begin{equation}
    \mathcal{L} \phi=\lim _{t \rightarrow 0} \frac{1}{t}\left[ \mathcal{K} ^t- Id \right] \phi.
\end{equation}
If $\mathbf{x}_t$ follows a SDE, $\mathcal{L}$ becomes a second-order linear differential operator. Specifically, for overdamped Langevin dynamics with potential energy $E$ and at temperature $T$,
\begin{equation}
    \mathrm{d} \mathbf{x}_t=-\nabla E\left(\mathbf{x}_t\right) \, \mathrm{d} t+\sqrt{2 k T} \,\mathrm{d} W_t,
\end{equation}
the corresponding generator is
\begin{equation}
    \mathcal{L} \phi(x)=-\nabla E \cdot \nabla \phi+k T \Delta \phi,
\end{equation}
where $\nabla \phi$ is the gradient and $\Delta \phi$ is the Laplacian. 

In essence, gEDMD uses statistical samples from the equilibrium distribution, or invariant measure, associated with some stochastic dynamics, to estimate an infinitesimal time-continuous generator of the dynamics. 

Given a basis set $\left\{\phi_i\right\}_{i=1}^n$ and data $\left\{\mathbf{x}_l\right\}_{l=1}^m$ sampled from a probability distribution, for example the Boltzmann distribution $\mu$, the finite-dimensional estimate for the generator 
$\mathcal{L}$ is represented as a matrix $\mathbf{L}$, which can be calculated from the solution to a system of linear equations
\begin{equation}
    \mathbf{L} = \mathbf{G}^{-1} \mathbf{A}.
\end{equation}
As the sample size $m \rightarrow \infty$, the matrices $\mathbf{G}$ and $\mathbf{A}$ can be expressed as expectations:
\begin{equation}
    \mathbf{G} \to \mathbb{E}[\phi(\mathbf{x}) \otimes \phi(\mathbf{x})], \quad \mathbf{A} \to \mathbb{E}[\phi(\mathbf{x}) \otimes \mathcal{L} \phi(\mathbf{x})].
\end{equation}
Here, $\phi(\mathbf{x})=\left[\phi_1(\mathbf{x}),\,\hdots,\,\phi_n(\mathbf{x})\right]$ are the vectors of basis functions evaluated at $\mathbf{x}$, and $\,\mathcal{L} \phi(\mathbf{x})$ is the corresponding generator applied to the basis functions.
Using a finite sample of data, we approximate $\mathbf{G}$ and $\mathbf{A}$ with the empirical averages
\begin{equation}
    \quad\mathbf{G}=\frac{1}{m}\sum_{l=1}^m\phi\left(\mathbf{x}_l\right)\otimes \phi\left(\mathbf{x}_l\right), \quad \text{and} \quad\mathbf{A}=\frac{1}{m} \sum_{l=1}^m \phi\left(\mathbf{x}_l\right)\otimes\mathcal{L}\phi\left(\mathbf{x}_l\right).
\end{equation}
For reversible overdamped Langevin dynamics, the expression for \(\mathbf{A}\) simplifies to 
\begin{equation}
\label{eq:gEDMD_rev_estimator}
    \mathbf{A}^* = -kT \mathbb{E}_{\mathbf{x}\sim \mu}[\nabla \phi(\mathbf{x}) \cdot \nabla \phi(\mathbf{x})],
\end{equation}
where $\nabla \phi(\mathbf{x})$ is the Jacobian matrix of the basis set $\phi$. The reversible estimator~\eqref{eq:gEDMD_rev_estimator} retains the symmetry and positive-definiteness of the generator matrix $\mathbf{A}$. The key implication is that we can estimate the generator $\mathcal{L}$ directly from equilibrium samples drawn from the Boltzmann distribution $\mu$ corresponding to the stationary state of the generator. The eigenvalues of this generator correspond to the kinetic transition rates between metastable states, providing a bound on the system’s slow kinetics. This approach allows us to approximate slow processes and rare transitions without requiring time-correlated trajectories, making it an efficient method for studying molecular kinetics based purely on equilibrium data.

\section{Results}
We evaluate the performance of the ambient and latent TI methods by applying them to study three different systems: a one-dimensional asymmetric double-well potential (see the Supporting information, ADW dataset generation, for dataset details) and MD simulation data \cite{diez2024generation} of two molecules from the QM9 dataset \cite{ramakrishnan2014quantum}: N-Methylformamide (N-Me) and 3-propan-2-ylhex-1-yne (3p2y1y). We use the learned maps to generate low-temperature conformational ensembles from high-temperature ensembles, compute free energy differences, and estimate the temperature dependence of kinetic exchange rates. Specific hyperparameter choices for all our experiments can be found in the Supporting information (hyperparameters for ADW system and hyperparameters for molecular systems).

\subsection{Generating Low Temperature Samples From High-Temperature Ensembles}
\begin{figure}
    \centering
    \includegraphics[width=0.8\linewidth]{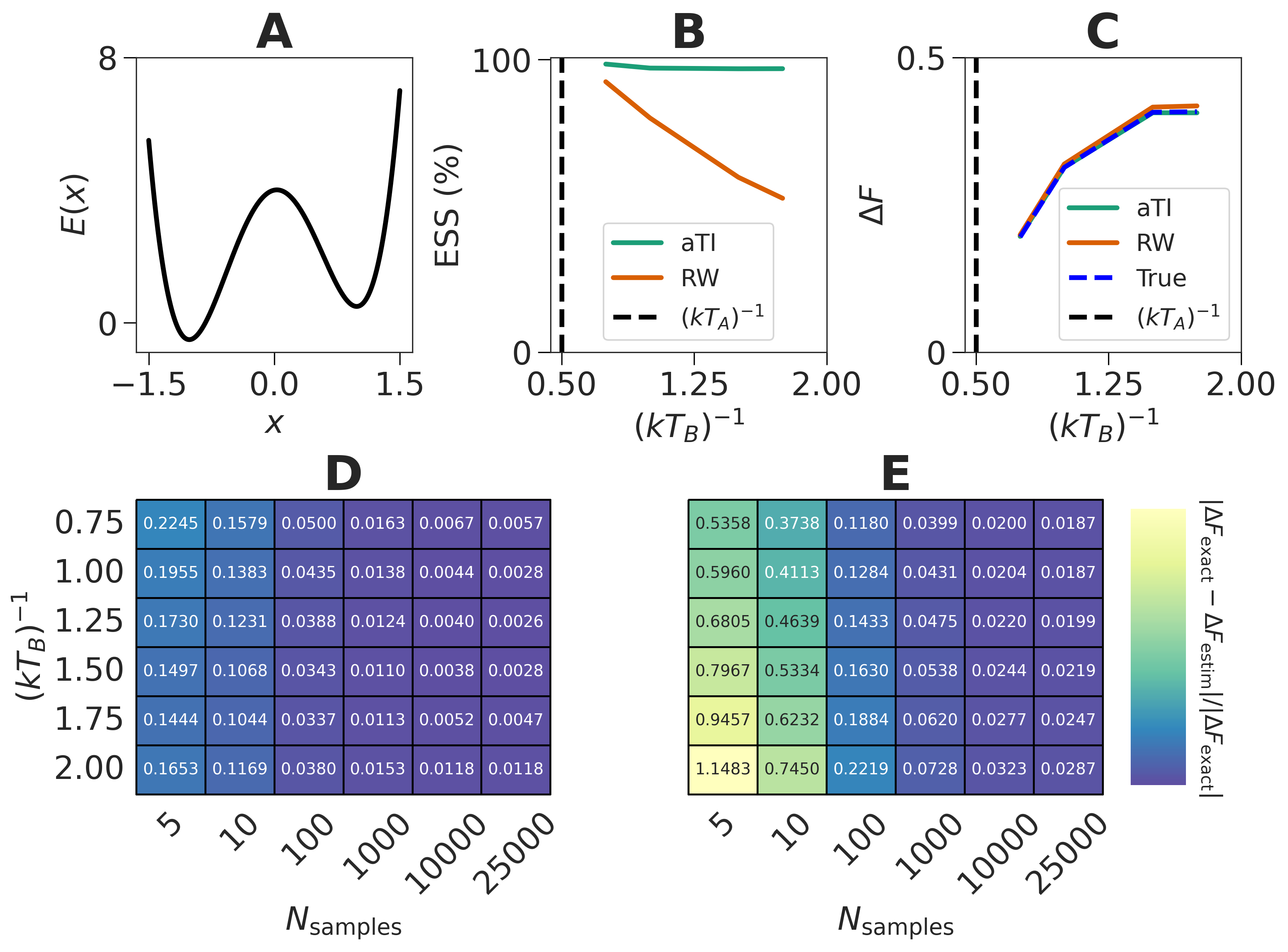}
    \caption{{\bf Results for Asymmetric double well potential.}\textbf{A} The energy landscape of the 1D asymmetric double well model system. \textbf{B} Effective Sample Sizes (ESS) for the aTI model compared to direct reweighing. Here, the free energy was estimated using $\Delta F^\mathrm{(TFEP)}$. \textbf{C} Estimated differences in Helmholtz free energy $\Delta F$ for the ambient TI (aTI) model and using direct re-weighing compared to true reference values.  \textbf{D}--\textbf{E} Heatmaps of relative errors in $\Delta F$, for the aTI model and direct re-weighing baseline respectively, as function of the number of samples used in the estimator and the transformation target $(kT_B)^{-1}$. In \textbf{B}--\textbf{E}, all transformations are made from $(kT_A)^{-1}=0.5$, where the aTI model was trained on data at $(kT_\mathrm{train})^{-1}\in\{0.5,\,1.25,\,2.0\}$. \textbf{A-E} To avoid high variance in the free energy estimators we made use of a filtering strategy discussed further in the Supporting information (IQR-filtering of outliers).}
    \label{fig:ti_interpolation_main_result_adw}
\end{figure}

We perform the first evaluation on a 1D asymmetric double well system (Fig.~ \ref{fig:ti_interpolation_main_result_adw}A). Here we train an aTI model on MD generated samples at temperatures $(kT_\mathrm{train})^{-1}\in\{0.5,\,1.25,\,2.0\}$ and transform from $(kT_A)^{-1}=0.5$ into different target temperatures not seen by the model during training. We achieve high sampling efficiency, measured through the Kish effective sample size (ESS) \cite{kishDesignEffect} (Fig.~ \ref{fig:ti_interpolation_main_result_adw}B). The ESS estimates the effective number of statistically independent samples, as
\begin{equation}
    n_{\mathrm{eff}} = \frac{(\sum_i w_i )^2}{\sum_i (w_i^2)},
\end{equation}
where $w_i$ is the importance weight of sample $x_i$ generated from a surrogate.

\begin{figure}[h!]
    \centering
    \includegraphics[width=0.8\linewidth]{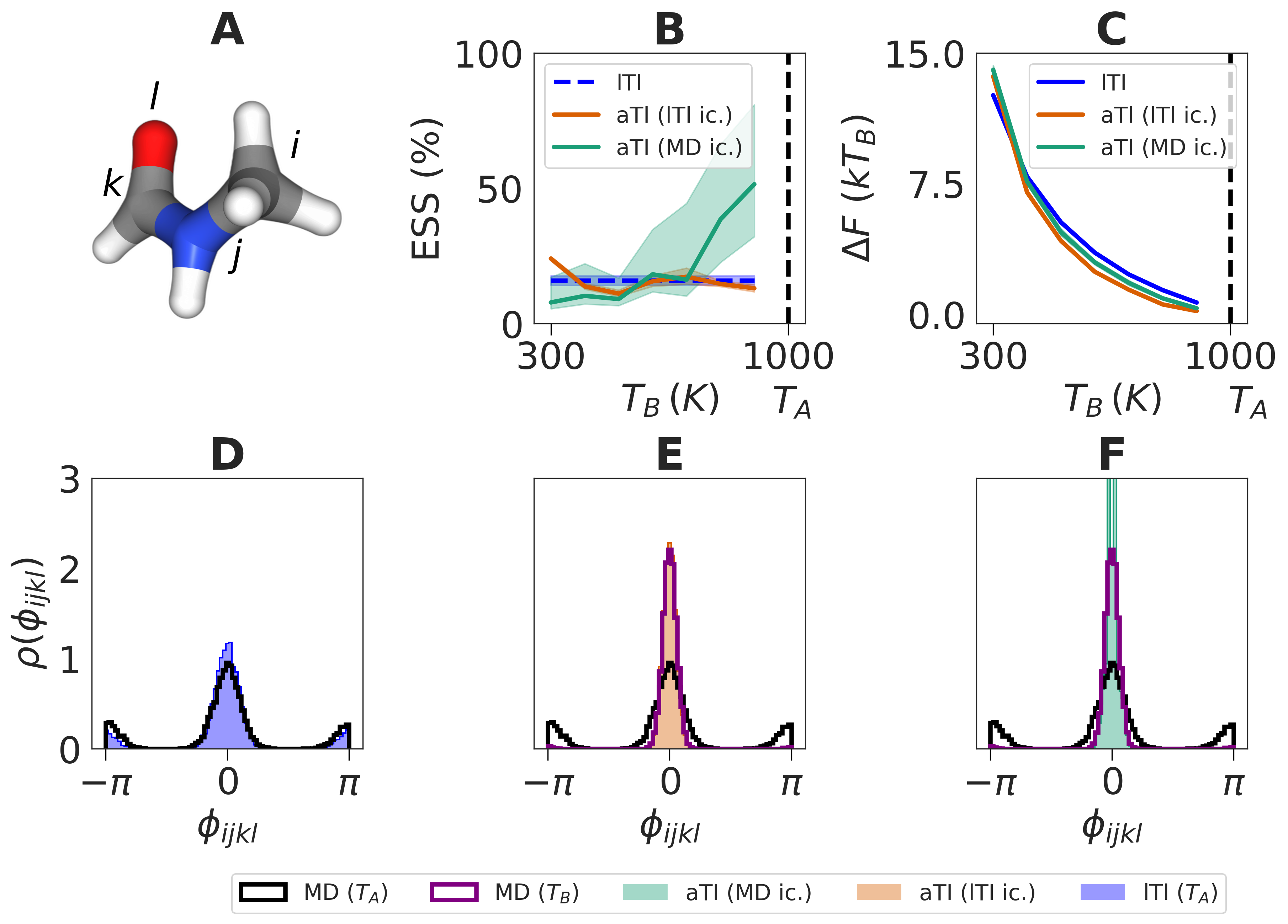}
    \caption{{\bf N-Me ensemble, effective sample sizes, and free energies.}\textbf{A} A visualization of the N-Me molecule with labels for the atoms $i,\,j,\,k,\,l$ which form the torsion angle $\phi_{ijkl}$. \textbf{B} The Effective Sample Size (ESS) plotted against the target temperature $T_B$. \textbf{C} Estimated differences in Helmholtz Free Energy $\Delta F$ between temperatures $T_A=1000\,\mathrm{K}$ and $T_B\in\{300,\,400,\,500,\,600,\,700,\,800,\,900\,\mathrm{K}\}$ for the latent TI (lTI), and ambient TI (aTI) applied to lTI and MD-simulated initial conditions. Here, the free energy was estimated using $\Delta F^\mathrm{(TFEP)}$. \textbf{D}--\textbf{F} Marginal histograms of the torsion angle between the four atoms $i,\,j,\,k$ and $l$, depicted in the small molecule. \textbf{D} Marginal histograms of torsions corresponding to lTI samples and MD data at $T_A=1000\,\mathrm{K}$. \textbf{E} Results of applying aTI to initial conditions generated through lTI, so that the system temperature is lowered from $1000\mathrm{K}$ to $300\,\mathrm{K}$, compared to reference MD simulated data. \textbf{F} Results of applying aTI to MD simulated initial conditions, so that the system temperature is lowered from $1000\,\mathrm{K}$ to $300\,\mathrm{K}$, compared to reference MD simulated data. \textbf{B-E} To avoid high variance on the free energy estimators we made use of a filtering strategy discussed further in the Supporting information (IQR-filtering of outliers)}
    \label{fig:ti_main_result_mdqm9}
\end{figure}

\begin{figure}[h!]
    \centering
    \includegraphics[width=0.8\linewidth]{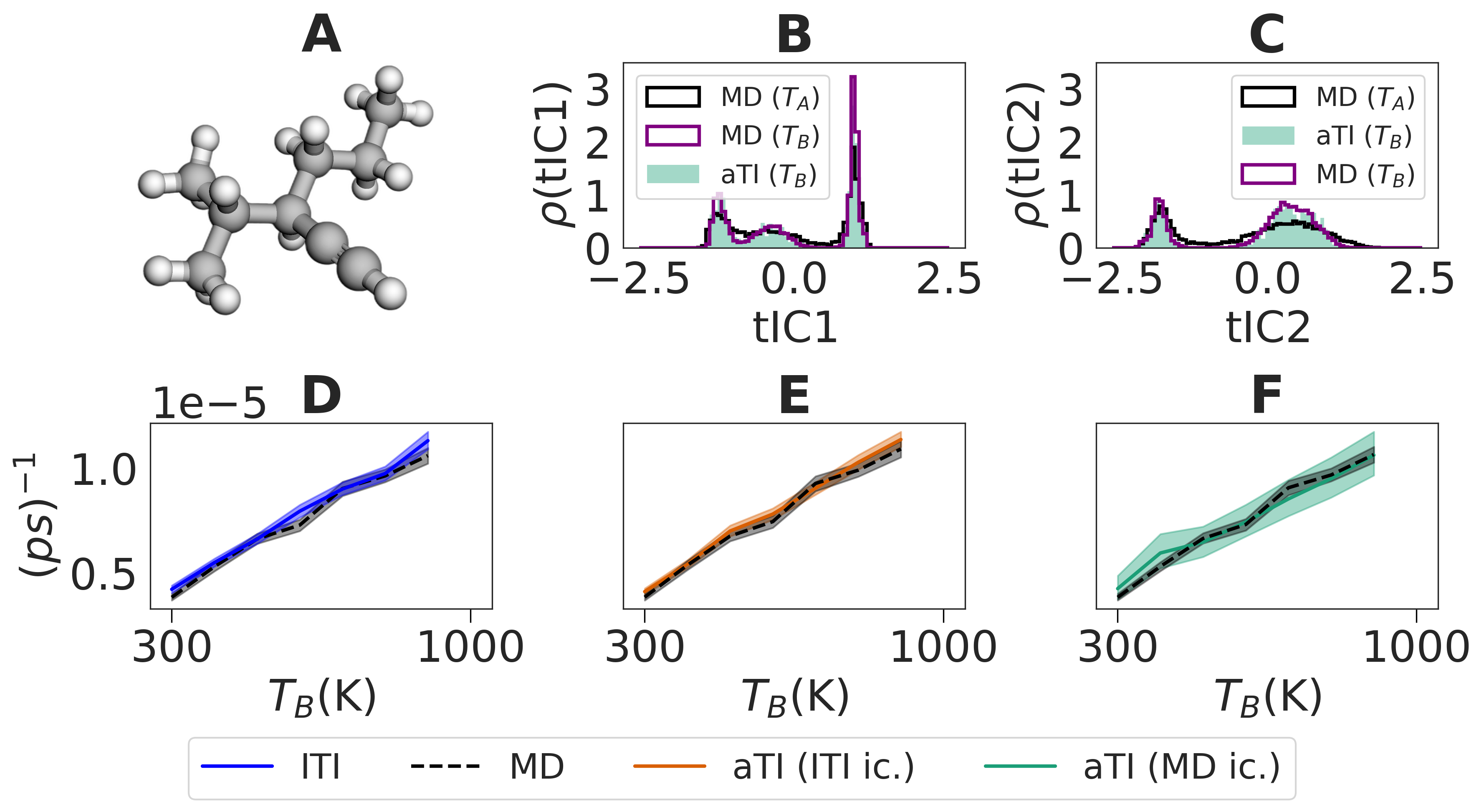} 
    \caption{{\bf 3p2y1y ensemble and kinetics without reweighing.} \textbf{A} A visualization of the 3p2y1y molecule. \textbf{B} The aTI torsion angles projected onto the first TICA ($\tau=2\,\mathrm{ps}$) dimension $\mathrm{tIC1}$, plotted in a histogram. Along with the torsion angles, we also show reference MD values at the initial and target temperatures $T_A$ and $T_B$. \textbf{C} The aTI torsion angles projected onto the second TICA dimension $\mathrm{tIC2}$, plotted in a histogram. Along with the torsion angles, we also show reference MD values at the initial and target temperatures $T_A$ and $T_B$. \textbf{D-F} Kinetic rate results estimated using gEDMD method for the larger molecule system, plotted against different target temperatures $T_B\in\{300,\,400,\,500,\,600,\,700,\,800,\,900\mathrm{K}\}$ for the lTI, and the aTI approaches, compared with MD samples.}
    \label{fig:tica_big_mol}
\end{figure}

We extend this approach to the molecular systems N-Me and 3p2y1p, training molecule-specific aTI models on the replica-exchange molecular dynamics from two molecules from the MDQM9 dataset \cite{diez2024generation}. At high temperatures, N-Me displays the multimodal distribution of a torsion (visualized in Fig.~\ref{fig:ti_main_result_mdqm9}A), which collapses into a unimodal distribution at low temperatures (Fig.~\ref{fig:ti_main_result_mdqm9}D-F). For 3p2y1y, two meta-stable states in the high temperature ensemble split into three distinct states transitioning from high to low temperature ensembles, as visualized by time-lagged independent components \cite{PhysRevLett.72.3634,Ziehe1998, PrezHernndez2013} (TICA) (Fig.~\ref{fig:tica_big_mol}) (see details in the Supporting information, scaling to larger systems). 

Using our aTI models we generate low-temperature ensembles, starting from high-temperature samples.  We generate these low-temperature ensembles in two ways: first, we transform samples  generated with conventional MD simulations at high temperature. Second, we generate samples using a BG-type  surrogate of the high-temperature ensemble, $T_A=1000\,K$, here a lTI model. We compare these results against an unbiased low-temperature MD simulations (Fig~\ref{fig:ti_main_result_mdqm9}D-F). To test aTIs ability to generalize beyond seen data, we train seven aTI models, such that each one is `blind' to the target temperature $T_B$ during training, i.e. $T_\mathrm{train}\in\{300,\,400,\,500,\,600,\,700,\,800,\,900,\,1000\,\mathrm{K}\}\setminus T_B$. In this way, we either have an interpolation or an extrapolation at test-time. We find that aTI models accurately generalizes a transformation from the initial Boltzmann distribution to a different thermodynamic state characterized by the target temperature $T_B$, indicated by the match of the histograms along the torsion angles Fig~\ref{fig:ti_main_result_mdqm9}D-F) and high Kish \cite{kishDesignEffect} ESS (Fig~\ref{fig:ti_main_result_mdqm9}B). We note that these ESS are lower for the higher dimensional molecular system compared to the 1D model potential, due to the increased complexity in the map, and improving these results are expected in line with the continued improvement in training strategies. Scaling up to the larger molecule 3p2y1y, we qualitatively find very similar results with distributions in the reduced TICA coordinates closely matching the reference simulations (Fig.~\ref{fig:tica_big_mol}B-C).

Finally, we find that training a single model that learns using multiple temperatures, provide much better predictions across the temperature range than specialized lTI and aTI models (Supporting information, single vs. multi-temperature training, Figure S3). This observation suggests that our TI models learn to share information between different thermodynamic ensembles, and thereby be applicable in cases with only limited data akin to what has been reported for multi-ensemble Markov models \cite{tram}.

\subsection{Predicting Free Energies Changes Upon Temperature Change}
Since CNFs allow for exact evaluation of changes in sample probability, we can use any of the free energy estimators discussed above to estimate the free energy difference between states $A$ and $B$. Using the obtained low-temperature samples and their corresponding changes in probability, we estimate free energy differences between the thermodynamic state at the initial and target temperatures with the TFEP estimator $\Delta F^\mathrm{(TFEP)}$.

For the ADW system, we compare our free energy estimates against an estimate based on numerical integration of the Boltzmann distributions which acts as a highly accurate benchmark. We further compare against a benchmark where we directly reweigh samples from the reference state $A$ to $B$ using importance sampling. Both aTI and RW accurately reproduce the numerical benchmark (Fig.~\ref{fig:ti_interpolation_main_result_adw}B), however, even for this simple system the ESS drops rapidly with the difference in $T_A$ and $T_B$, whereas the aTI based approach maintains a near perfect sample efficiency (Fig.~\ref{fig:ti_interpolation_main_result_adw}C).

To illustrate the practical impact of the ESS on free-energy estimates, we compute the relative errors in free energy as a function of samples from the reference state $A$ for aTI (Fig.~\ref{fig:ti_interpolation_main_result_adw}D) and direct reweighing (Fig.~\ref{fig:ti_interpolation_main_result_adw}E). Unsurprisingly, we find that we can get highly accurate estimates with as little as five samples using aTI whereas comparable errors would need 20-fold more samples in the direct reweighing case. These results further underline, the potential of TI models as a data-efficient way to learn maps between thermodynamic states and compute free energy differences. Moving on to the molecule N-Me we get consistent estimates of the free energy differences using lTI and aTI (Fig.~\ref{fig:ti_main_result_mdqm9}C). While the estimates of the aTI estimator using samples from a surrogate (lTI) and MD as initial condition give us comparable $\Delta F$ estimates, we note that the surrogate appears to limited the ESS (Fig.~\ref{fig:ti_main_result_mdqm9}B). For a comparison of the two free energy estimators $\Delta F^\mathrm{(TFEP)}$ and $\Delta F^\mathrm{(BG)}$, see the Supporting information (free energy perturbation methods, Figure S1), where we empirically find the bound (eq.~\ref{eqn:tfep_for_bg}) to hold, with a gap suggesting that the learned aTI and lTI maps are not perfect, in line with the lower ESS (Fig.~\ref{fig:ti_main_result_mdqm9}B).

\subsection{Generator Estimation to Approximate Molecular Kinetics Across Temperatures}
\begin{figure}[h!]
    \centering
    \includegraphics[width=0.8\linewidth]{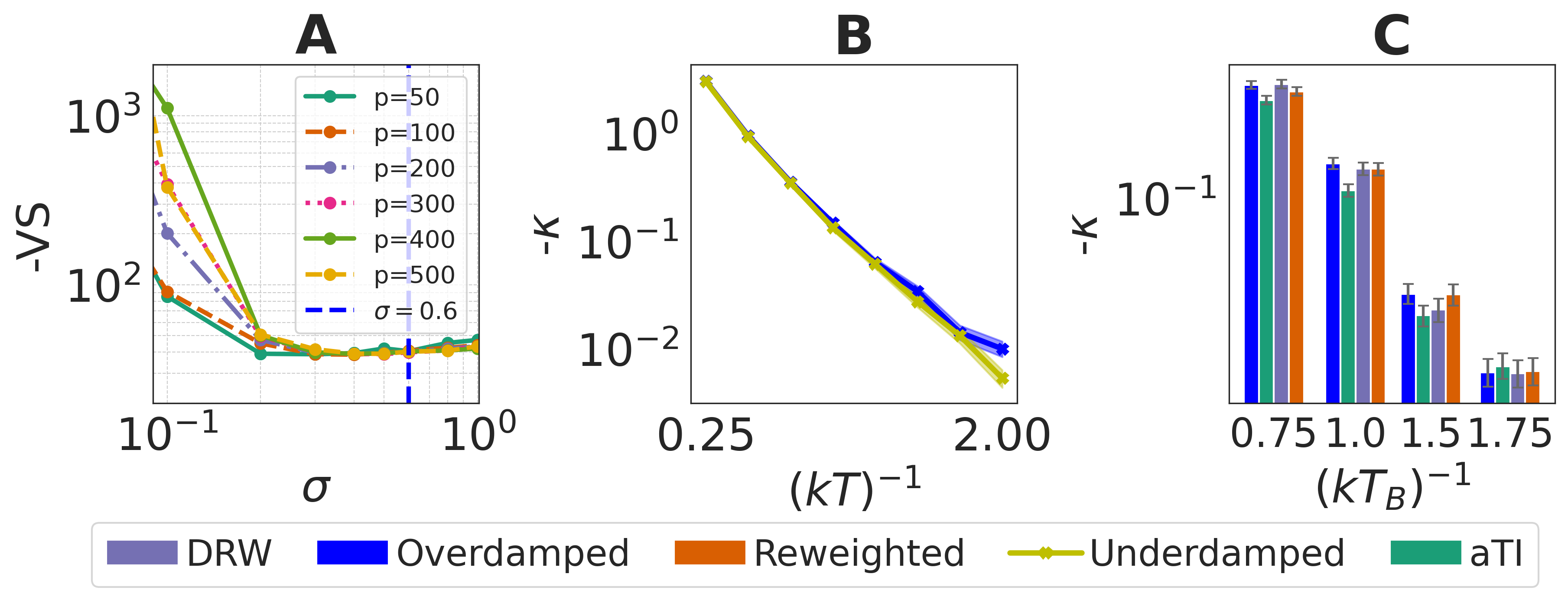}
    \caption{{\bf gEDMD analysis of Asymmtric Double Well system}
    \textbf{A} VAMP scores\cite{wu2020variational} as a function of the kernel bandwidth \(\sigma\) for different numbers of Fourier features \(p\). We select \(p = 50\) and \(\sigma = 0.6\) as the model parameters. 
    \textbf{B} Kinetic rates calculated under the Brownian assumption using gEDMD methods for both overdamped and underdamped MD samples.
    \textbf{C} Kinetic rates estimated using gEDMD for MD samples, aTI predictions, and reweighted results. The 
    aTI model was trained on data at \((kT_\mathrm{train})^{-1} \in \{0.5,\, 1.25,\, 2.0\}\). Transformations are made from \((kT_A)^{-1} = 0.5\) to \((kT_B)^{-1} \in \{0.75,\, 1.0,\, 1.5,\, 1.75\}\).
    }
    \label{fig:gedmd_adw}
\end{figure}

Next, we leverage our ability to efficiently generate unbiased statistics across a temperature range to show how kinetics change across a temperature range. Following previous work, we use a kernel-based gEDMD estimator~\cite{klus_kernel-based_2020} with Random Fourier Features (RFFs)~\cite{rahimi2007random, nuske2023efficient}. We are particularly interested in characterizing the temperature dependence of slow processes, corresponding to the smallest eigenvalues of the generator, which correspond directly to relaxation rates associated with exchange between meta-stable states. The estimation of the gEDMD models with RFFs requires optimizing hyperparameters, including the bandwidth and number of features, as illustrated in (Fig.~\ref{fig:gedmd_adw}A). Details of this model selection process are provided in the Supporting information (gEDMD model selection).

\begin{figure}[h!]
    \centering
    \includegraphics[width=0.8\linewidth]{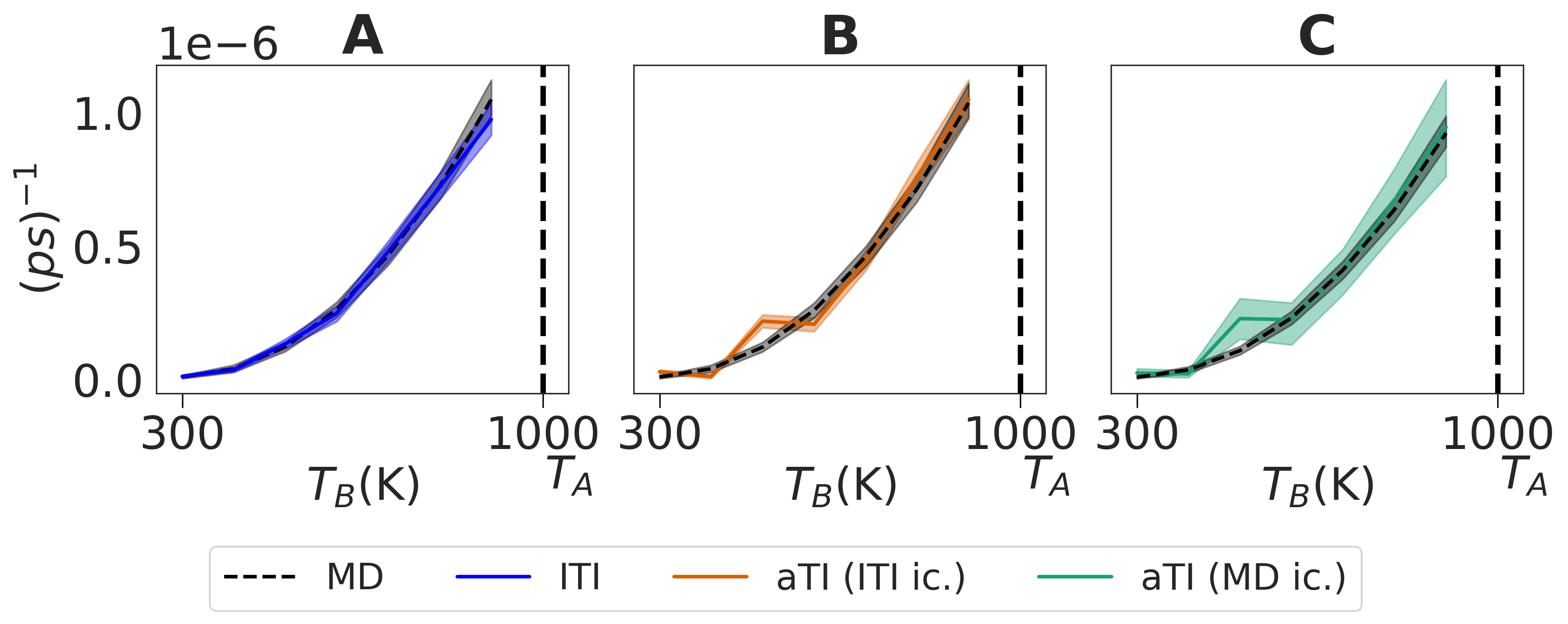}
    \caption{{\bf Kinetic rate results estimated using gEDMD for N-Me}, plotted against different target temperatures $T_B\in\{300,\,400,\,500,\,600,\,700,\,800,\,900\mathrm{K}\}$ for (A) the latent TI (lTI), and the two ambient TI (aTI) approaches with initial conditions generated using lTI  (B) and MD (C) . We show rates computed using MD samples with a dashed line. The reference temperature $T_A$ is shown with a vertical dashed line.}
    \label{fig:qm9_gedmd}
\end{figure}

 For ADW, we focus on the slowest rate, which is associated with exchange between the two major states (Fig.~\ref{fig:ti_interpolation_main_result_adw}A). Using gEDMD we compute these rates across four temperatures comparing aTI samples with and without reweighing using importance weights, to direct reweighing and overdamped Langevin simulations. Broadly, the predicted rates all align with reweighted samples yielding most closely following the overdamped reference (Fig.~\ref{fig:gedmd_adw}B-C). Since using overdamped Langevin (Brownian) dynamics is uncommon on molecular applications as assumed with the estimator of the generator, we provide rates exacted from MSMs trained on simulations conducted with overdamped and underdamped simulations illustrating that rates extracted from overdamped simulations bound the corresponding rates from underdamped simulations from below (Supporting information, MSM for different systems). For the molecular systems, we compute kinetic rates in a similar fashion: assuming the generator is Brownian and using a torsion angle as features resolving the main meta-stabilities of the systems. We compare the lTI, and aTI applied to the lTI and MD initial conditions, illustrating that the predicted temperature dependence of the kinetic rates are consistent with reference values computed from MD samples, both for the N-Me molecule (Fig.~\ref{fig:qm9_gedmd}C) and the 3p2y1y molecule (Fig.~\ref{fig:tica_big_mol}D-F). Strikingly, we observed that although our model imperfectly learns the map between thermodynamic states the predicted rates do not differ dramatically if we reweigh samples to the exact target (Figs.~\ref{fig:tica_big_mol}D-F and ~\ref{fig:gedmd_adw}C), suggesting that gEDMD is robust to some modeling bias introduced by generative surrogates.

\section{Conclusion}
In this work, we have introduced TI as a flexible approach to equilibrium sampling of Boltzmann distributions across different thermodynamic states, and as a practical approach for free energy estimation through TFEP and for analysing the kinetic dependence of thermodynamic transformations. We present two different instances of TI, one a directly mapping between states in the configurational space --- ambient TI --- and an alternative approach mapping between thermodynamic states through a latent reference state --- latent TI. We audition TI using transformations between different temperatures, and find both lTI and aTI models show promise of efficient generation of statistics from ensembles at multiple different temperatures and accurately estimate changes in free energy. Further, we find that the model interpolates and extrapolates to temperatures not seen during training. Due to the high data and sampling efficiency of TI we envision its use to overcome slow sampling at low temperatures, for example through integration with Parallel Tempering schemes akin to recent work \cite{skipping_re_ladder}. Combining TI with gEDMD we can study the temperature dependence of kinetics on thermodynamic transformations. We illustrate this through predicting the temperature dependence of exchange between two meta-stable states in the molecule N-Me and by analysing the slowest process in the 3p2y1y molecule where we recover qualitative kinetics using biased samples. 

A major bottleneck for scalable deployment of TI depends on the specific application. The calculation of change in log-probability relies on the computation of Jacobian determinants of a velocity field during sampling, which scales linearly with the dimension of the system and the number of integration time-steps. Consequently, for larger systems where transformations are more complicated and dimensions are higher, exact computation of free energies is not tractable in the current architectures. However, there is a steady development in the machine learning community to make architectural as well as algorithmic improvements to speed up such calculations, by enforcing structure in the Jacobian or improving conditioning of the learned velocity fields using ideas from optimal transport. Both of these ideas will be necessary to ensure the competitiveness of TI and related approaches compared to current approaches. In the meantime, as we illustrate, our proposed TI approaches still yield semi-quantitative predictions for larger systems, even if reweighing is expensive. Further, future work might benefit from coarse-graining to reduce the number of particles. However, here explicitly dealing with non-Markovian dynamics might be necessary to obtain accurate models \cite{Cao2023}. As such, we believe that TI is an important step towards learnable transformations between thermodynamic states, enabling calculations of free energy and kinetics. 

\begin{acknowledgement}
This work was partially supported by the Wallenberg AI, Autonomous Systems and Software Program (WASP) funded by the Knut and Alice Wallenberg Foundation. We further acknowledge funding from the Knut and Alice Wallenberg Foundation project: "From atom to organism: Bridging the scales in the design of ion channel drugs." Preliminary results were enabled by resources provided by the National Academic Infrastructure for Supercomputing in Sweden (NAISS) at Alvis
(project: NAISS 2024/22-33), partially funded by the Swedish Research Council through grant
agreement no. 2022-06725. 

The authors thank Juan Viguera Diez for sharing MDQM9-nc dataset ahead of publication and help with evaluation and Ross Irwin and Johann Flemming Gloy for helpful discussions.
\end{acknowledgement}

\bibliography{ref}

\begin{tocentry}
    \includegraphics[width=\linewidth]{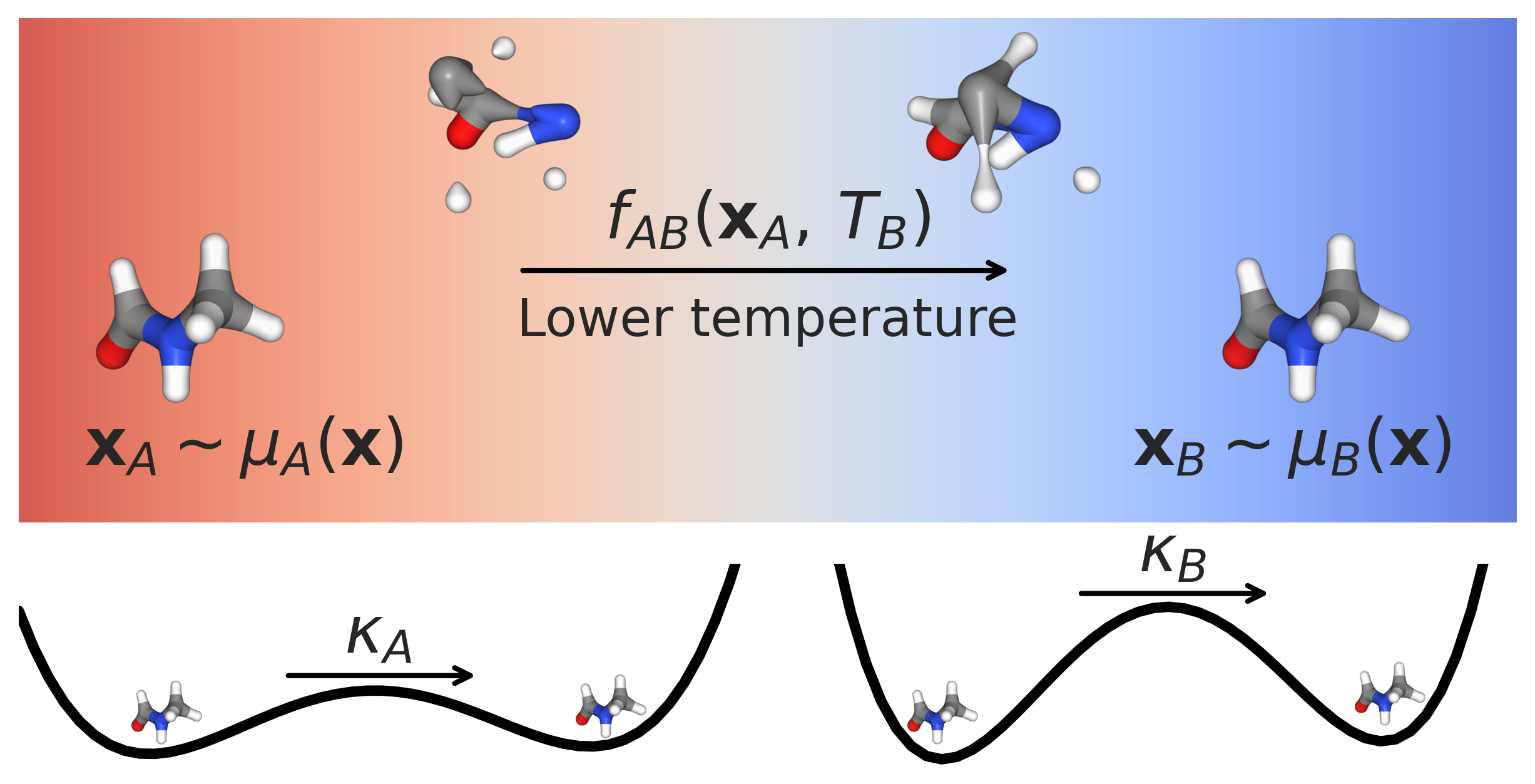}
\end{tocentry}

\end{document}